# Empirical characteristics of different types of pedestrian streams

Jun Zhang[a,*], Armin Seyfried[a,b]

[a]*Computer Simulations for Fire Safety and Pedestrian Traffic, Wuppertal University, Pauluskirchstrasse 11, 42285 Wuppertal, Germany*
[b]*Jülich Supercomputing Centre, Forschungszentrum Jülich GmbH, D-52425 Jülich, Germany*

**Abstract**

Reliable empirical data and proper understanding of pedestrian dynamics are necessary for fire safety design. However, specifications and data in different handbooks as well as experimental studies differ considerably. In this study, series of experiments under laboratory conditions were carried out to study the characteristics of uni- and bidirectional pedestrian streams in straight corridor. The Voronoi method is used to resolve the fine structure of the resulting velocity–density relations and spatial dependence of the measurements. The result shows differences in the shape of the relation for $\rho > 1.0$ m$^{-2}$. The maximal specific flow of unidirectional streams is significantly larger than that of all bidirectional streams examined.

**Nomenclature**

| | |
|---|---|
| $v$ | velocity (m/s) |
| $\rho$ | density (m$^{-2}$) |
| $J_s$ | specific flow ((m·s)$^{-1}$) |
| $J$ | flow (1/s) |
| $A_i$ | size of the *i*th Voronoi cell (m$^2$) |
| $A_m$ | size of the measurement area (m$^2$) |
| $b$ | width of the corridor (m) |
| $C_{sb}$ | capacity of short bottleneck ((m·s)$^{-1}$) |
| $C_{lb}$ | capacity of long bottleneck ((m·s)$^{-1}$) |
| $C_{cor}$ | capacity of straight corridor ((m·s)$^{-1}$) |

## 1. Introduction

At present, the fire safety design on building facility can mainly be divided into two categories: prescriptive method and performance-based method [1]. Using a prescriptive method means the facility should be designed to meet the minimum safety requirement which is standardized in the form of rules in legal code for design of building, e.g., minimum width or length of corridor. With the development of modern building technologies, materials and design concept, the constructions, shapes and functions of buildings change a lot. In this situation, prescriptive methods becomes insufficient to dimension escape routes for more complicated buildings such as indoor arenas, shopping malls, or underground railway stations. As a response to architects and designers who want more flexibility, performance-based methods which are performed according to handbooks and simulation results have been developed. In this methods crowd movement is quantitatively specified using the fundamental diagram which gives the relationship between density, velocity and flow. Nearly all handbooks include fundamental diagrams for design of facilities. However, specifications of different experimental studies, guidelines and handbooks, display large discrepancies even for the most relevant characteristics like maximal flow values, the corresponding density $\rho(J_{s,m})$ and the density where the flow is expected to become zero due to overcrowding [2]. As seen in Table 1, the basic variables are compared for the common used handbooks of Weidmann [3], Predtechenskii and Milinskii (PM) [4], Fruin [5] and SFPE handbook [6]. The fundamental diagrams for corridors, ramps, aisles and doorways are distinguished only in the handbook of PM and they are unified into one equation in other handbooks. Unlike other handbooks, the differences of fundamental diagrams between uni- and bidirectional pedestrian flow are considered in the handbook of Fruin.

---

* Corresponding author. Tel.: +49-202-439-4241
*E-mail address:* jun.zhang@uni-wuppertal.de

Table 1. Comparison of the main design variables in different handbooks.

| | Weidmann [3] | PM* [4] | Fruin [5] | SFPE [6] |
|---|---|---|---|---|
| Maximal density $\rho_m$ [1/m$^2$] | 5.4 | 8.14 | 5.4 | 3.8 |
| Free velocity $v_0$ [m/s] | 1.34 | 0.95 | 1.27 | 1.19 |
| $J_{s,m}$ [(m·s)$^{-1}$] | 1.22 | 1.49 | 1.43 | 1.30 |
| $\rho(J_{s,m})$ [1/m$^2$] | 1.75 | 6.64 | 2.16 | 1.88 |
| $v(J_{s,m})$ [m/s] | 0.7 | 0.2 | 0.7 | 0.7 |
| Distinction between corridor and opening | No | Yes | No | No |
| Distinction between uni- and multi-directional flow | No | - | Yes | No |

* Note that, the average dimension 0.113 m$^2$ per person is used for transforming the units of variables of PM.

However, up to now there is no consensus whether the fundamental diagrams for uni- and bidirectional flows differ from each other or not. Weidmann [3] and Predtechenskii and Milinksii [4] neglected the differences in accordance with Fruin, who stated that the fundamental diagrams of multi- and uni-directional flow differ only slightly [5]. This disagrees with results of Navin and Wheeler [7] who found a reduction of the flow in dependence of directional imbalances. Pushkarev et al. [8] and Lam et al. [9, 10] assume that bidirectional flows are not substantially different from unidirectional flow as long as the densities of the opposite streams are not too different. However, Older et al. stated that different ratios of flows in bidirectional stream do not show any consistent effect on the walking speed [11]. Besides, Helbing et al. [12] concluded that counterflows are significantly more efficient than unidirectional flows. However, they compare average flow values without considering the influence of the density. Kretz et al. [13] have reported similar findings, but the influence of density and variations in time on the flow are not considered in this study. Figure 1 shows the fundamental diagrams of unidirectional [4, 14-18] and bidirectional flow [7, 11, 19-23]. It seems that the fundamental diagrams of unidirectional flow lie above those of bidirectional flow, especially for $\rho > 1.0$ m$^{-2}$. The actual characteristics and differences between them are not clear and need further analysis.

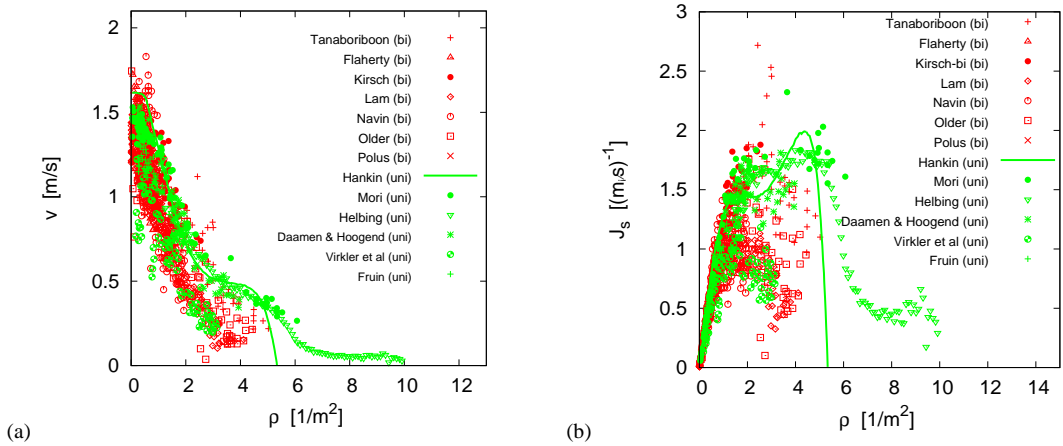

Fig. 1. Fundamental diagrams of uni- and bidirectional pedestrian flow from various studies. (a) Density–velocity and (b) density–specific flow.

The above discussion shows that up to now there is no consensus on the characteristics of different types of pedestrian flow. A lot of possible factors may influence the properties of pedestrian movement. To reduce uncontrollable influences as much as possible we used a homogenous group of test persons and carried out series of well-controlled laboratory experiments of pedestrian streams. In this study, we mainly investigate the fundamental diagrams of uni- and bidirectional pedestrian flow. We also study the dependence of pedestrian flows on different kind of geometries.

The structure of this paper is as follows. In section 2 we describe the setup of the experiment. Section 3 describes the measurement method, data extraction and some main results. Finally, the conclusions from our investigations will be discussed.

## 2. Experiment Setup

To minimize uncontrollable factors and to compare the fundamental diagrams of different types of flow, well-controlled laboratory experiments were carried out with the same group of test persons in hall 2 of the fairground Düsseldorf (Germany) in May 2009. These experiments include pedestrian movement in a straight corridor, closed ring, T-junction as well as around a corner. Up to 400 persons composed mostly of students participated in the experiments. The mean age and height of the participants were 25 ± 5.7 years old and 1.76 ± 0.09 m respectively. The free velocity $v_0 = 1.55 \pm 0.18$ m/s was obtained by measuring 42 participants' free movement.

All runs of the experiments were recorded by two synchronized stereo cameras of type Bumblebee XB3 (manufactured by Point Grey). They were mounted on the rack of the ceiling 784 cm above the floor with the viewing direction perpendicular to the floor. The cameras have a resolution of 1280 × 960 pixels and a frame rate of 16 fps (corresponding to 0.0625 second per frame). To increase the region of observation, the left and the right part of the scenario were recorded by the two cameras separately. The overlapping field of view of the stereo system is $\alpha = 64°$ at the average head distance of about 6 m from the cameras. With the above mentioned height range, all pedestrians can be seen without occlusion at any time. The geometrical variations of the boundaries and moving directions of pedestrians in each experiment will be introduced in the following sections.

Figure 2 shows a sketch of the experiment setup and a snapshot of the experiments to study unidirectional flow in a straight corridor with open boundaries. Three straight corridors with the widths of 1.8 m, 2.4 m and 3.0 m were chosen. To regulate the pedestrian density in corridor, the widths of the entrance $b_{entrance}$ and the exit $b_{exit}$ were changed in each run. At the beginning of each run, the participants were held within a waiting area behind the entrance and then passed through a 4 m passage into the corridor. The passage was used as a buffer to minimize the effect of the entrance. In this way, the pedestrian flow in the corridor was nearly homogeneous over its entire width. When a pedestrian leaves through the exit, he or she returned to the waiting area for the next run. 28 runs of unidirectional pedestrian experiments were carried out in all.

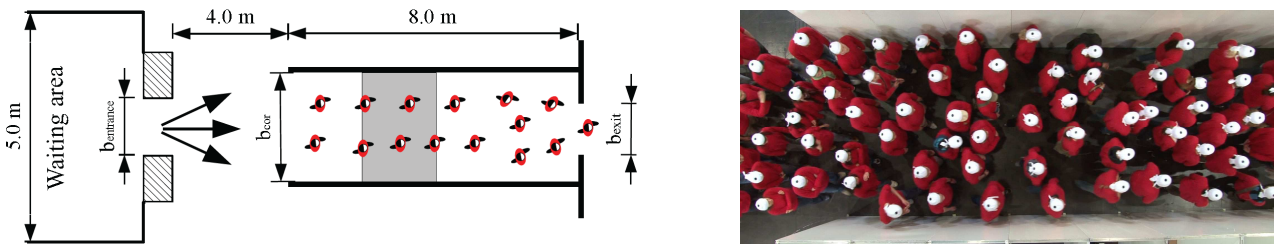

Fig. 2. Setup and snapshot of unidirectional flow experiment. The gray area in the sketch shows the location of measurement area.

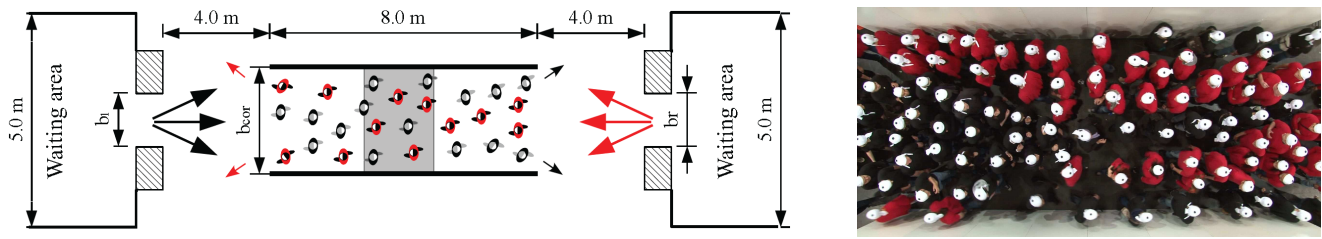

Fig. 3. Setup and snapshot of bidirectional flow experiment. The gray area in the sketch shows the location of measurement area. Dynamical lane formation can be observed from the snapshot.

Figure 3 shows the sketch of the experiment setup and a snapshot of the experiment to study bidirectional flow in a straight corridor. Two groups of participants located in the waiting areas at the left and right side of corridor and then entered the corridor through the entrances when a run started. The width of the left entrance $b_l$ and the right entrance $b_r$ were changed in each run to regulate the density in the corridor and the ratio of the opposing streams. Altogether 22 runs of bidirectional pedestrian streams were performed in corridors with widths of 3.0 m and 3.6 m respectively. Unlike with the unidirectional flow experiment, we gave different instructions to the participants on which exit to choose. One is asking the test persons to choose the exit freely, the other one is letting them choose the exit according to a number given to them in advance. The persons with odd numbers should choose the left exit in the end, while the ones with even number were asked to choose the exit in the right side. With freely choosing of exits the stable separated lanes (SSL) are observed, whereas the dynamical multi lanes (DML) are obtained by specifying the exits in advance.

## 3. Analysis and Results

*3.1. Trajectories*

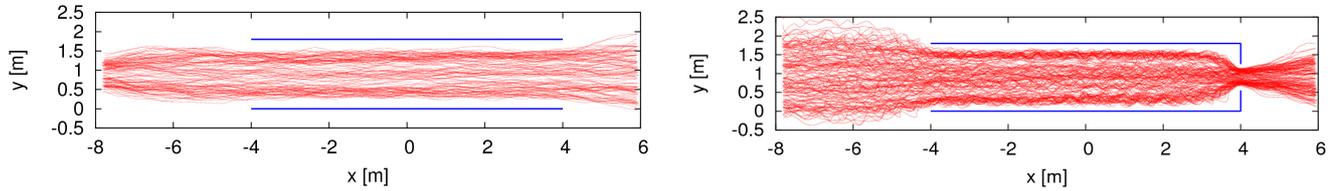

(a) Unidirectional pedestrian flow at different densities.

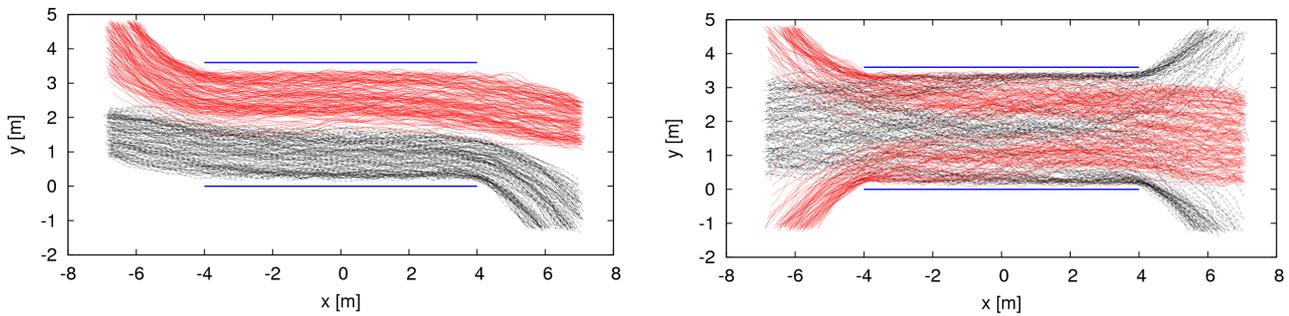

(b) Different kinds of bidirectional pedestrian flow.

Fig. 4. Trajectories of the pedestrians in four runs of the experiments. The data are extracted from the video recordings by *Petrack*. The formation of lanes can be observed from the trajectories for all these runs. For the bidirectional flow experiment the trajectories of right and left moving pedestrian are represented in different colors. The distances between the edge of trajectories and the boundary are also not the same in various density situations.

To make a highly precise analysis, accurate pedestrian trajectories were automatically extracted from video recordings using the software *PeTrack* [24]. Lens distortion and perspective view are taken into account in this program. Figure 4(a) shows the pedestrian trajectories for two runs of unidirectional flow at low and high density situations. To obtain higher crowd density, a bottleneck at the end of the corridor was formed. The middle lane is wider at high density situation, whereas the distance between the edge of trajectories and the boundary is smaller than that at low density situation. Figure 4(b) displays the pedestrian paths for bidirectional flow with stable separated lanes and dynamical multi lanes respectively. From these trajectories, pedestrian characteristics including flow, density, velocity and individual distances at any time and position can be determined.

*3.2. Measurement Method*

In reference [25] and [26], the effect of different measurement methods on the fundamental diagram of pedestrian flow has been compared. Minor influences are found for the density ranges observed in the experiment but only the Voronoi method is able to resolve the fine structure of the fundamental diagram. Consequently, in this study we focus on the Voronoi method for its high spatial resolution in combination with low fluctuation.

This method is based on Voronoi diagrams [27] which are a special kind of decomposition of a metric space determined by distances to a specified set of objects in the space. To each such object one associates a corresponding Voronoi cell. The distance from the set of all points in the Voronoi cell to the given object is not greater than their distance to the other objects. At any time the positions of the pedestrians can be represented as a set of objects, from which the Voronoi diagrams are generated. The cell area $A_i$ can be thought as the personal space belonging to each pedestrian $i$. Then, the density and velocity distribution of the space $\rho_{xy}$ and $v_{xy}$ (see Figure 5) are defined as

$$\rho_{xy} = 1/A_i \text{ and } v_{xy} = v_i(t) \text{ if } (x, y) \in A_i \tag{1}$$

Where $v_i(t)$ is the instantaneous velocity of each person at time $t$ (see [24]).
The Voronoi density and velocity for the measurement area $A_m$ is defined as

$$<\rho>_v (x,y,t) = \frac{\iint \rho_{xy} dxdy}{A_m}, \tag{2}$$

$$<v>_v (x,y,t) = \frac{\iint v_{xy} dxdy}{A_m}. \tag{3}$$

The specific flow

$$<J_s>_v (x,y,t) = <\rho>_v (x,y,t) \cdot <v>_v (x,y,t) \tag{4}$$

is calculated using the Voronoi density and velocity according to the hydrodynamic relation $J = \rho v b$. $J_s$ represent the flow rate of pedestrians passing through unit width of facility. It is necessary for the quantification and comparison of capacities of different kinds of facilities.

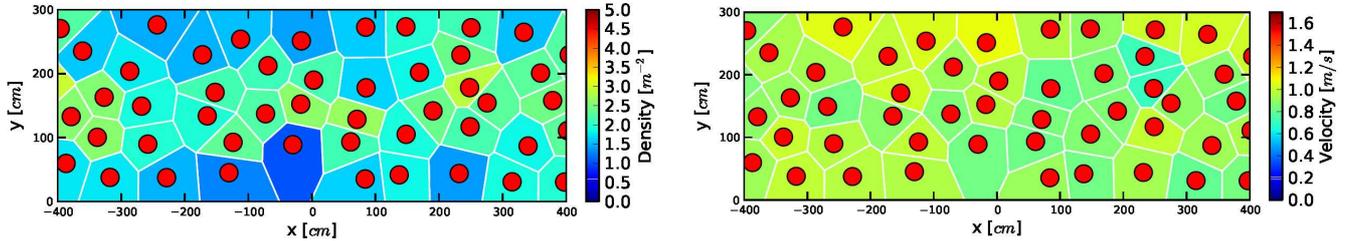

Fig. 5. Distribution of density and velocity over space obtained from Voronoi method for unidirectional flow at a certain time. Note that the Voronoi diagrams are generated according to the spatial distribution of pedestrians which are represented with red circles.

*3.3. Results*

According to the study in [25] and [28], the ordering in bidirectional flow has little influence on the fundamental diagram and the specific concept works well for both uni- and bidirectional flow at the density ranges in our experiments. Thus, we are free to choose the experiments with a larger range of density to compare their characteristics. Here we select the runs in 3.0 m width corridor for unidirectional flow and in 3.6 m width corridor for bidirectional flow with dynamical multi lanes to make analysis. One of the remarkable things is that the data of the unidirectional flow for $\rho > 2.0$ m$^{-2}$ are obtained by slight change of the experiment setup. This may limit the comparability of fundamental diagrams for $\rho > 2.0$ m$^{-2}$.

Figure 6(a) shows the relationship between density and velocity for uni- and bidirectional pedestrian flow. At densities of $\rho < 1.0$ m$^{-2}$, no significant difference exists. For $\rho > 1.0$ m$^{-2}$, however, the velocities for unidirectional flows are larger than that of bidirectional flows. The difference between the two cases becomes more apparent in the flow-density diagram (Figure 6(b)) where a qualitative difference can be observed. In the bidirectional case a plateau is formed starting at a density $\rho \approx 1.0$ m$^{-2}$ and the flow becomes almost independent of the density. Such plateaus are typical for systems which contain "defects" which limit the flow and have been observed e.g. on bidirectional ant trails [29] where they are a consequence of the interaction of the ants. In our experiments the defects are conflicts of persons moving in the opposite direction. These conflicts only happen between two persons but the reduction of the velocity influences those following.

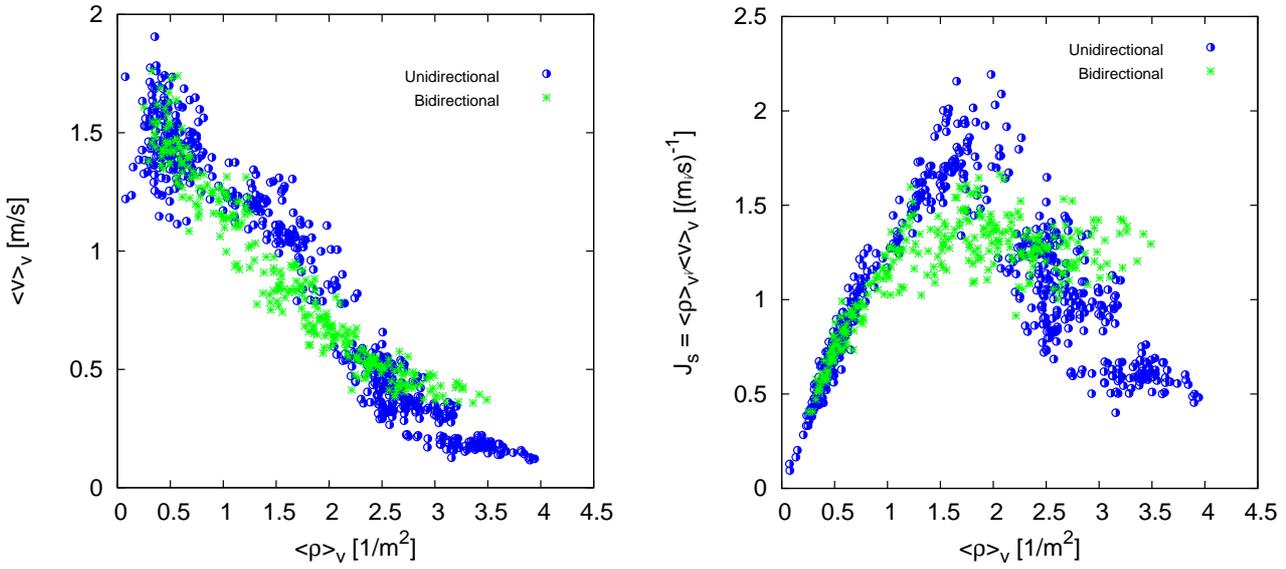

Fig. 6. Comparison of the fundamental diagrams between uni- and bidirectional pedestrian flow. Note that the data for unidirectional flow is obtained from the experiment in 3 m wide corridor, whereas the other is for the bidirectional flow with dynamical multi lanes in 3.6 m wide corridor. The trend of the relations and maximum specific flow are totally different for these two types of flows especially for $\rho > 1.0$ m$^{-2}$.

This difference in the fundamental diagrams implies that pedestrian flow with stable separated lanes should not be interpreted as two unidirectional flows. Although the self-organized lanes can decrease the head-on conflicts, interactions between the opposing streams are still relevant.

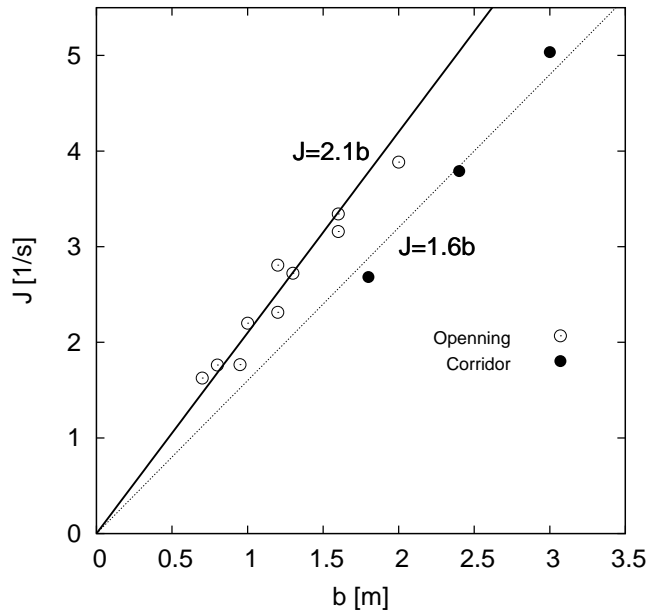

Fig. 7. Influence of the width of facility on the flow. The open points are obtained from the runs for $b_{cor} > b_{exit}$, while the solid points are the maximum flow rate in corridor observed in the experiment of unidirectional flow. Linear relations between the flow rate and facility width are observed, however, the slopes of the lines are various for different types of facilities.

We further study the influence of the geometries on the flows. With the current setting of the experiment for unidirectional flow, we can investigate the relationship between the flow rate $J$ and the opening width $b_{exit}$ as well as the corridor width $b_{cor}$. In Figure 7 it shows nearly linear relationships $J = 2.1 \cdot b$ for opening (short bottleneck *sb*) and $J = 1.6 \cdot b$ for the corridor. Here the flow rate is calculated by directly counting the number of pedestrians passing a reference line in geometries. Note that the solid points in the figure are the maximum flow rates through the corridor (at the location $x = 0$) that have been observed in the experiments. The open points are obtained at the exit (at $x = 4$) from the runs with $b_{cor} > b_{exit}$. From the result it seems that the specific flow through the opening is 2.1 (m·s)$^{-1}$ when congestions occur in front of the opening. The same relationship $J = 1.9 \cdot b$ has also been obtained for the long bottleneck (*lb*) in [30]. If all the maximum values of flow we obtained from our experiments are the capacity $C$ of the facilities, we could conclude that it should be $C_{sb} > C_{lb} > C_{cor}$. The short bottlenecks have higher flow values compared to long bottlenecks. This finding agrees with that in [31]. However, it does not agree with most handbooks like SFPE which do not distinguish between the fundamental diagrams of these three types of facilities. However, we could not check whether the capacity of the opening is larger than that of the corridor or not, because with current experiment setup we do not know whether higher specific flow can be observed in the condition of $b_{cor} = b_{exit}$.

## 4. Summary

A series of well-controlled laboratory pedestrian experiments of uni- and bidirectional flow were performed in straight corridors with up to 400 persons. The whole processes of the experiment were recorded using two synchronized stereo cameras. The trajectories of each pedestrian are extracted with high accuracy from the video recordings automatically using *PeTrack*. We use the Voronoi method in this study to analyze the data of the experiments for its small fluctuation and high resolution in time and space. The fundamental diagrams of uni- and directional flow show clear differences in the experiment. The maximum flow value is about 2.0 (m·s)$^{-1}$ for unidirectional flow while 1.5 (m·s)$^{-1}$ for bidirectional flow. This should be considered during the design of facilities where uni- and bidirectional streams would appear. The self-organized lanes can help to relief the head-on conflicts effectively and increase the ordering of the stream. However, these conflicts do not affect the fundamental diagram of bidirectional flow. The properties of pedestrian flow through different types of geometries including corridor and bottlenecks are also compared. Short bottlenecks have higher flow values compared to long bottlenecks, while the straight corridor has the highest flow. In the design of emergency exits, the influence of the shape and geometry of facilities on the pedestrian flow rate should be considered. For different types of geometries, the selected parameters should not keep constant all the time.